\documentstyle[12pt]{article}

\renewcommand{\thefootnote}{\fnsymbol{footnote}}
\newcommand{\newsection}{
\setcounter{equation}{0}
\section}

\def\appendix#1{
  \addtocounter{section}{1}
  \setcounter{equation}{0}
  \renewcommand{\thesection}{\Alph{section}}
  \section*{Appendix \thesection\protect\indent \parbox[t]{11.715cm} {#1}}
  \addcontentsline{toc}{section}{Appendix \thesection\ \ \ #1}
  }

\def\e{{\,\rm e}\,}
\newcommand{\rf}[1]{(\ref{#1})}
\newcommand{\eq}[1]{Eq.~(\ref{#1})}
\newcommand{\non}{\nonumber \\*}

\def\be{\begin{equation}}
\def\ee{\end{equation}}
\def\bea{\begin{eqnarray}}
\def\eea{\end{eqnarray}}

\def\d{\partial}

\def\E{F}

\newcommand{\fr}[2]{{\textstyle {#1 \over #2}}}

\newcommand{\beq}{\begin{eqnarray}}
\newcommand{\eeq}{\end{eqnarray}}

\begin{document}
\begin{titlepage}
\begin{flushright}
NBI--HE--99--18\\
ITEP--TH--26/99\\
hep-th/9906134\\
June 1999
\end{flushright}

\begin{center}

{\Large\bf Screening and D-brane Dynamics\\[.4cm] in Finite Temperature
Superstring theory}

\vspace{1cm}

{\bf J. Ambj\o rn}, {\bf Yu.\ Makeenko}\footnote{Permanent Address: 
Institute of
Theoretical and Experimental Physics, B. Cheremushkinskaya 25, 117218 Moscow,
Russia.}, {\bf G.W. Semenoff}\footnote{On
leave from Department of Physics and Astronomy,
University of British Columbia, 6224 Agricultural Road, Vancouver,
British Columbia, V6T 1Z1 Canada.} and {\bf R.J. Szabo}\\
\vspace{24pt}
{\it The Niels Bohr Institute\\
Blegdamsvej 17, DK-2100\\
Copenhagen {\O}, Denmark\\}\vskip .4cm
E-mail: {\tt ambjorn,makeenko,semenoff,szabo@nbi.dk}

\end{center}

\vskip 1 cm
\begin{abstract}

The thermal dynamics of D-branes and of open superstrings in background gauge
fields is studied. It is shown that D-brane dynamics forbids
constant velocity motion at finite temperature. T-duality is used to
interpret this feature as a consequence of the absence of an equilibrium state
of charged strings at finite temperature in a constant background electric
field, as a result of Debye screening of electric
fields. The effective action for the Polyakov loop operator is computed and the
corresponding screening solutions are described. The finite temperature theory
is also used to illustrate the importance of carefully incorporating Wu--Yang
terms into the string path integral for compact target spaces.

\end{abstract}

\end{titlepage}
\setcounter{page}{2}
\renewcommand{\thefootnote}{\arabic{footnote}}
\setcounter{footnote}{0}

\newsection{Introduction and Summary}

One of the fundamental problems in superstring theory is to understand the
behaviour of strings in external fields. It is relevant to question about the
existence and nature of possible non-perturbative vacuum states. Using duality,
it is also important for understanding various aspects of D-brane dynamics. An
example of an external field problem which has been studied extensively is that
with a constant electromagnetic field \cite{ft}--\cite{Tse98}. Gauge fields
couple to charges which live at the endpoints of open strings, and the
modification of the vacuum energy by a slowly varying gauge field is known to
be described by the Born--Infeld action \cite{ft}
\begin{equation}
S_{\rm BI}=-\frac{1}{(2\pi)^9(\alpha')^5 g_s}\int d^{10}x~{\rm Tr}\left\{
\sqrt{-\det_{\mu,\nu}\left(\eta_{\mu\nu}+2\pi e\alpha' F_{\mu\nu}
\right)}+\dots\right\}
\label{BI}\end{equation}
where the factor of inverse string coupling comes from the fact that this
is a tree level (disc) amplitude. Here and in the following $\dots$ will denote
higher order terms in the (covariant) derivative expansion of the external
fields. There are corrections to the effective action \rf{BI} at the one-loop
(annulus diagram) and higher loop orders. By T-duality, the Born--Infeld action
gives the effective dynamics of D-branes \cite{Bac96} and is very useful in
determining the couplings of degrees of freedom in the D-brane to string theory
as well as supergravity degrees of freedom. In this paper we will study the 
effective action for gauge fields in a thermal
state of superstring theory. We will be particularly interested in how the
Born--Infeld action \rf{BI} is modified by temperature and what this
modification
implies about electric fields. We shall compute the leading temperature
corrections to (\ref{BI}) for weakly-coupled strings and slowly-varying
external fields.

In the path integral approach to finite temperature superstring theory,
the spacetime is taken as ten dimensional Euclidean space with time $x^0$
compactified on a circle of circumference $\beta=1/k_{\rm B}T$, where $k_{\rm
B}$ is Boltzmann's constant and $T$ is the temperature.  Then,
Euclidean external gauge fields must be periodic in $x^0$,
\be
A_\mu(x^0,\vec x)= A_\mu(x^0+\beta,\vec x)
\ee
In this case, it is impossible to fix a gauge where $A_0=0$. On the other hand,
it is possible to fix a gauge where $A_0$ is independent of the Euclidean time
$x^0$ and is diagonal, $(A_0)_{bc}(x^0,\vec x)=a(\vec x)^b\,\delta_{bc}$. We
will consider the special case where the external gauge field is in an Abelian
$U(1)$ subgroup of the full non-Abelian Chan--Paton gauge group, and set all
spatial components of the gauge fields to 0. The generator of the Abelian
subgroup is characterized by the eigenvalues $e_b$ which are interpreted as
$U(1)$ charges and
\be
A_0(\vec x)=\left( \matrix{ e_1 & 0 & 0 & \dots\cr
                            0 & e_2 & 0 & \dots\cr
                            0 & 0 & e_3 & \dots\cr
                            \dots&~&~&\dots\cr}\right) ~ a(\vec x)
\ee

The disc amplitude (\ref{BI}) is unmodified at finite temperature, 
because the disc worldsheet cannot wrap the cylindrical target
space and thus cannot distinguish between a compactified and an
un-compactified spacetime. The first corrections due to temperature appear in
the annulus amplitude. We shall find that the free energy is given by the
functional
\begin{equation}
F[a]=\int d^{9}x~\left\{
\frac{1}{(2\pi)^7(\alpha')^{5}g_s}\sum_{b}\sqrt{1+\left(
2\pi \alpha'e_b\vec\nabla a(\vec x)\right)^2}+
\sum_{b,c}V_{\rm eff}\Bigl[(e_b-e_c)a(\vec x)\Bigr]+\dots\right\}
\end{equation}
where, for the type I NSR superstring, the effective potential is
\be
V_{\rm eff}\left[z\right]=\frac{32\pi}{(2\pi^2\alpha')^5}\,
\int\limits_0^\infty\frac{dt}{t^6}~\Theta_2\left(\frac\beta\pi\,
z\left|\frac{i\beta^2}{2\pi^2\alpha't}\right.\right)
\prod_{n=1}^\infty\left(\frac{1+\e^{-2\pi nt}}{1-\e^{-2\pi nt}}\right)^8
\ee
and $\Theta_\alpha$ denote the standard Jacobi theta-functions.

The dependence of the effective potential on the temporal component of the
gauge field is familiar from finite temperature gauge theory. In particular,
it implies Debye screening of electric fields. Consider the linearized equation
for the minima of the free energy, which after rescaling can be written in
the form
\begin{equation}
-{\vec\nabla}^2 a(\vec x)+\mu^2 a(\vec x)=0
\end{equation}
This equation has exponentially decaying solutions $a(\vec x)\sim\e^{-\mu\vert
\vec x\vert}$ where, at low temperatures ($\sqrt{\alpha'} k_{\rm B}T\ll1$), the
Debye screening mass $\mu$ is given by
\be
\mu^2=3\pi^2\cdot2^{22}\,
g_s(\alpha')^3(k_{\rm B}T)^8~\frac{\sum_{b,c}(e_b-e_c)^2}{\sum_be_b^2}
+O\left(\e^{-1/\sqrt{\alpha'}k_{\rm B}T}\right)
\label{debyemass}\ee
It is clear for this reason that constant electric fields cannot be
extrema of the effective action. Another way to see that the existence
of constant electric fields is incompatible with the existence of a Debye mass
is the following. If we try to make a constant Abelian electric field by
choosing the background gauge field
\be
a(\vec x)=\vec E\cdot\vec x
\ee
then the integral over $\vec x$ that one would have to do in computing the
temperature corrections to the free
energy would vanish because of the property
\begin{equation}
\int d^9x~\Theta_2\left(\frac\beta\pi\,
\vec E\cdot\vec x\left|\frac{i\beta^2}{2\pi^2\alpha't}\right.\right)=0
\end{equation}
which is a consequence of supersymmetry. This property can be interpreted as a
result of Debye screening, i.e. that the finite temperature string theory
forbids constant electric fields. All states except the ground state contain
excitations of charged particles and therefore have infinite energy. In fact,
because of the Schwinger mechanism \cite{BP92}, even the ground state is
unstable.

In the following sections we will demonstrate this result directly using the
boundary state formalism. Two different gauges are commonly used to study
constant electric fields, the static gauge, $(A_0,\vec A)=(\vec E\cdot\vec
x,0)$, and the temporal gauge, $(A_0,\vec A)=(0,-\vec Ex^0)$. In the temporal
gauge, the gauge potential is not periodic in Euclidean time, but it is
periodic up to a gauge transformation,
\be
\vec A(x^0+\beta,\vec x)=\vec A(x^0,\vec x)+\vec\nabla\left(
-\beta\vec E\cdot \vec x\right)
\label{gaugetransftemp}\ee
In this case, it is necessary to augment the usual coupling of the edge of a
charged open string to the gauge field by adding a Wu--Yang term \cite{WY75},
$\oint A_\mu\,dx^\mu\rightarrow \oint A_\mu\,dx^\mu +\chi$, in order to
compensate the gauge transformation \rf{gaugetransftemp}. The geometrical
reasoning for adding this term is explained in Appendix A. In the next section
we shall use the example of a relativistic charged particle to illustrate
how the Wu--Yang term is essential if the partition function is to be
independent of the gauge choice. One byproduct of the following
analysis will therefore be the importance of incorporating such terms into the
superstring path integral for generic target space compactifications
involving external fields.

Under T-duality, external gauge fields map onto the trajectories of D-branes,
whose long wavelength dynamics are described by supergravity. Thermal states of
D-branes are of interest as non-BPS states of superstring theory. In the
supergravity picture, they have a natural Hawking temperature and radiation
which can be interpreted as the emission of closed string modes by a
non-extremal D-brane configuration. It has been suggested \cite{susskind} that
the gravitational Hawking temperature and the temperature of a Boltzmann gas of
D-branes should be identified. It is therefore natural to study the
thermodynamics of D-branes and to understand the
special aspects of D-brane dynamics which arise when they are in a thermal
state. It has been shown \cite{AMS98} that the effective action for a gas
of D0-branes which is obtained by summing over physical (GSO projected)
superstring states is
\begin{equation}
S_{\rm eff}[\vec x]=\int\limits_0^\beta d\tau~\left\{
\frac{1}{\sqrt{\alpha'}g_s}\,
\sqrt{1+\dot {\vec
x}(\tau)^2}-\frac{256}{\beta}\,\e^{-2\beta\vert\vec
x\vert/2\pi\alpha'}+\dots\right\}
\label{SeffD0}\end{equation}
The first term, which is the relativistic free particle Lagrangian for the
D0-branes, follows from the Born--Infeld action \rf{BI} and T-duality. It is
correct up to higher derivatives of the relative position vector $\vec x(\tau)$
with respect to the Euclidean time $\tau$. The second term, which we have given
the low temperature limit of, comes from the annulus amplitude and is valid
only for static branes. Corrections to the annulus amplitude at finite
temperature which take into account the
time dependence of $\vec x$ are not known.

Unlike at zero temperature, where the system of static D0-branes is a BPS state
and this potential would vanish due to supersymmetry, temperature breaks
supersymmetry and leaves the residual short-ranged attractive interaction. The
potential in \eq{SeffD0} was obtained from the superstring annulus free energy
with Dirichlet boundary conditions~\cite{Gre92}
\begin{equation}
F[\vec x,\beta,a_0]=
\frac{8}{\pi\sqrt{2\pi\alpha'}}\int\limits_0^\infty
\frac{dt}{t^{3/2}}~\e^{-{\vec x}^{\,2}t/2\pi\alpha'}\,
\Theta_2\left( \frac{\beta}{\pi}\,a_0\left|
\frac{i\beta^2}{2\pi^2\alpha't}\right.\right)
\prod_{n=1}^\infty \left( \frac{1+\e^{-2\pi nt}}{1-\e^{-2\pi nt}}\right)^8
\label{Tpot}\end{equation}
and incorporating the dynamics of the time component of the gauge field
$a_0(\tau)$ on the D-particle worldline. On the other hand, at zero temperature
the motion of an assembly of D0-branes at constant relative velocity $\vec
v=\dot{\vec x}$ also breaks supersymmetry. The leading velocity dependent
gravitational attraction between a pair of D0-branes in ten dimensional
supergravity is proportional to $\vec v^{\,4}/|\vec x|^7$ \cite{Bac96,DKPS}.
The exact one-loop potential at zero temperature is again given by the annulus
amplitude which in this case is \cite{Bac96,GG96,BCD}
\be
F[\vec x,\vec
v]=\frac1{\sqrt{2\pi\alpha'}}\int\limits_0^\infty\frac{dt}t~
\e^{-b^2t/2\pi\alpha'}\,\frac{\Theta_1\left(\left.\frac{\epsilon t}2
\right|it\right)^4}{\Theta_1(\epsilon t|it)}\,\left[\e^{-\pi t/12}\,
\prod_{n=1}^\infty\left(1-\e^{-2\pi nt}\right)\right]^{-9}
\label{freevelT0}\ee
where $b$ is the impact parameter for the scattering and $\epsilon$ is the
relative rapidity of the two branes. In the following we will describe the
situation when one tries to describe these two vacuum amplitudes collectively.

T-duality maps free open strings to open strings whose endpoints are attached
to D-branes. It is implemented in a straightforward way when the string action
depends only on worldsheet derivatives of a string embedding coordinate. It
then proceeds by replacing $\partial_a x^i$ by $i\epsilon_{ab} \partial_bx^i$
and the resulting replacement of Neumann boundary conditions for $x^i$ with
Dirichlet conditions. We are interested in examining the T-dual of the string
in a constant electric field $\vec E$. This should produce an open string whose
endpoints are constrained by Dirichlet boundary conditions to end on a D-brane
which is moving with a constant velocity $\vec v= 2\pi\alpha'e\vec E$
\cite{Bac96}. Indeed, in the temporal gauge, the coupling of the constant
electric field $\exp ie\oint \vec Ex^0\cdot \partial_{\rm t}\vec x$ is replaced
by the vertex operator for a moving D-brane $\exp-e\oint \vec E x^0\cdot
\partial_{\rm n} \vec x$ \cite{leigh}, where $\partial_{\rm t}$ and
$\partial_{\rm n}$ are
the derivatives tangential and normal to the boundary of the string worldsheet,
respectively. However, we shall find that, at finite temperature, both of these
couplings must be augmented by a Wu--Yang term which breaks the translation
invariance in the direction of the electric field $\vec E$. The T-dual of the
Wu--Yang term can be obtained in principle, but it is a complicated, non-local
expression. We shall find, however, that it has a simple presentation within
the boundary state formalism. Once it is taken into account, we can show that,
just as Debye screening forbids constant electric fields in open superstring
theory, it also forbids the constant motion of D-branes. This implies that
there is a damping of their motion analogous to Debye screening.

\newsection{Thermodynamic Partition Function in a Constant Electric Field}

\subsection{Relativistic Particle}

To illustrate the general ideas, it is instructive to begin with the case of a
relativistic charged scalar particle in a constant electric field at finite
temperature. The extension to strings will be straightforward and the final
result can be represented as a sum over particles with masses given by the open
string spectrum and appropriate degeneracies. The diagonal elements of the
(unnormalized)
thermal density matrix are given by the Euclidean path integral
\be
\rho(\vec y,\vec y;\beta)
= \frac 12 \int\limits_0^\infty \frac{ds}{s}~\e^{-\frac 12 M^2 s}\,
\int Dx^\mu(t)~\e^{-\frac 12 \int_0^s dt\,\dot x_\mu^2(t)}
\,\Phi[x^\mu(t)],
\label{rho}
\ee
where $\Phi[x^\mu(t)]$ is the Abelian phase factor in the given electromagnetic
background. The temperature $T=1/k_{\rm B}\beta$ enters via the boundary
conditions
\bea
x^0(s)&=&x^0(0)+n\beta ~~~(\hbox{integer}~n),\non
\vec x(s)&=&\vec x(0)~=~\vec y,
\label{pbc}
\eea
which compactifies the Euclidean time coordinate $x^0$ on a circle of
circumference $\beta$. The parameter $s$ plays the role of the proper time
during which the particle propagates along the given trajectory $x^\mu(t)$ that
can wind $n$ times around the space-time cylinder, while $M$ is the particle
mass. The path integral \rf{rho} is to be summed over all winding numbers
$n\in{\bf Z}$.

The constant electric field can be described in the static gauge by the
vector potential
\bea
A_0(x)&=&\frac{2\pi\nu}\beta- \vec \E\cdot \vec x,\non
\vec A(x)&=&\vec0 \;,
\label{gauge0}\eea
where we have denoted the ``temporal'' component $F_{0i}$ of the Euclidean
field strength tensor by $\vec {\E}\equiv\E_i (i=1,\ldots,d-1)$. It is related
to the electric field $\vec E$ in Minkowski space by
\be
\vec \E=i \vec E.
\label{FvsE}
\ee
The constant $\nu$ in \eq{gauge0} can always be taken to lie in the interval
$[0,1)$ due to the periodicity of $x^0$. It plays a crucial role at finite
temperature as we will discuss, but for now $\nu$ is simply associated with
choosing a reference point where the potential $A_0$ vanishes. Since $A_\mu$ is
single valued for the gauge choice~\rf{gauge0}, the phase factor is simply
\be
\Phi[x^\mu(t)]\equiv\exp i \int\limits_0^s dt~A_\mu\Bigl(x(t)\Bigr)\, \dot
x^\mu(t)=
\exp\left(2\pi i\nu n- i\int\limits_0^sdt~\vec \E\cdot\vec x(t)\,\dot
x^0(t)\right).
\label{phase0}
\ee

Note that one can choose a more general gauge
\bea
A_0( x)&=&\frac{2\pi\nu}\beta-(1-c)\vec \E\cdot\vec x,\non
\vec A(x)&=&c\vec \E x^0 ,
\label{gauge1}
\eea
so that~\rf{gauge0} corresponds to the choice $c=0$. For this choice, $\vec A$
is multivalued due to the boundary condition~\rf{pbc} so that the Wu--Yang
term~\cite{WY75} has to be included in the phase factor which now takes the
form
\bea
& &\Phi[x^\mu(t)]\nonumber\\& &=
\exp\left(2\pi i\nu n-i(1-c)\int\limits_0^s dt~\vec \E\cdot\vec x(t)\,\dot
x^0(t) +i c\int\limits_0^s dt~x^0(t)\,\vec \E\cdot\dot {\vec x}(t)
- icn\beta \vec \E\cdot\vec x(0)\right).\nonumber\\& &
\label{phase1}
\eea
This formula is derived in Appendix A. Integrating by parts, it is easy to see
that~\rf{phase0} and \rf{phase1} coincide as they should due to gauge
invariance.

Substituting~\rf{phase0} in \eq{rho}, we get a Gaussian path integral of the
form of that for the harmonic oscillator~\cite{Fey72}. It is convenient to
choose coordinates where $\vec \E=(\E,0,\ldots,0)$. The result then reads
\be
\rho(\vec y,\vec y;\beta)
= \frac \beta2 \int\limits_0^\infty \frac{ds}{s}~\e^{-\frac 12 M^2 s}\,
\frac{1}{(2\pi s)^{d/2-1 }} \frac{\E}{4\pi \sinh {\frac{\E s}{2}}}\,
\Theta_3 \left(\nu-\frac{\beta\vec \E\cdot\vec y}{2\pi}\left \vert
\frac{i\beta^2 \E}{4\pi \tanh \frac{\E s}2}\right.\right)
\label{finalrho}
\ee
where $\Theta_3$ is the usual Jacobi theta function
\be
\Theta_3 \left(\nu \, \vert i\tau\right) =
\sum_{n=-\infty}^\infty\e^{-\pi \tau n^2 + 2 \pi i \nu n }
\label{theta3}
\ee
and $d$ is the dimension of space-time. Eq.~\rf{finalrho} is similar to the
density matrix for (spinor) quantum electrodynamics at finite temperature in
the presence of a constant background electric field~\cite{LR92}. The path
integral derivation of it is sketched in Appendix B.

The formula~\rf{finalrho} resembles the density matrix for a harmonic
oscillator of frequency $\E$. This feature can be easily understood in second
quantization, where the eigenvalues of the operator $i\d_0$ which enters the
Klein--Gordon operator are given by the Matsubara frequencies $2\pi m /\beta$,
which results at each level $m$ in the Hamiltonian
\be
H_m=-\frac 12\,\vec\nabla\,{}^2 -\frac 12\Bigl(\d_0-i A_0\Bigr)^2 =
-\frac 12\,\vec\nabla\,{}^2 + \frac 12 \left(\frac {2\pi(m-\nu)} \beta + \E
x^1\right)^2
\ee
for the harmonic oscillator of frequency $\E$ oscillating along the axis 1 with
a shifted position of $x^1$. The density matrix is then given by
\be
\rho(\vec y,\vec y;\beta)
= \frac {1}2 \int\limits_0^\infty \frac{ds}{s}~\e^{-\frac 12 M^2 s}\,
\frac{1}{(2\pi s)^{d/2-1 }}\,
\sqrt{\frac{\E}{2\pi \sinh \E s}}\,
\sum_{m=-\infty}^\infty
\e^{-\E \tanh \frac{\E s}2
\left(y^1+\frac {2\pi (m-\nu)} {\beta \E}  \right)^2} \,.
\ee
After a Poisson resummation which is the statement that
\be
\Theta_3 (\nu | i\tau)= \frac 1{\sqrt\tau}\,\e^{-\pi \nu^2/\tau}\,
\Theta_3\left(\frac {i\nu}\tau\left| \frac i\tau\right.\right),
\ee
we get \eq{finalrho}.

The dependence on $\vec y$ in \eq{finalrho} implies
the violation of translational invariance of the theory.
Its appearance is most easily understood in
the temporal gauge (given by~\rf{gauge1} with $c=1$)
 where only $\vec A$ depends
on $x$. The term $-\beta \vec \E \cdot\vec y /2\pi$ in the
first argument of the theta function in~\rf{finalrho}
comes in this gauge entirely from the Wu--Yang term
\be
- n\beta \vec \E\cdot\vec x(0)=- n\beta \vec \E\cdot\vec y \,,
\ee
which was added to the exponent
of the phase factor $\Phi$ to guarantee global gauge invariance. 

If we were
instead to choose a vector potential which is a periodic function of $t=x^0$,
then the gauge field would be single-valued and no Wu--Yang term would be
required. The simplest example is a field $\vec A(t)$ which is a linear
function of $t$ for $0\leq t<\beta$ and then returns to its initial value
$\vec A(0)$ at $t=\beta$,
\be
\vec A_{\rm per}(t)= \vec \E t \Bigl( 1 -\theta(t-\beta) \Bigr),
\label{constantperiodic}
\ee
where $\theta$ is the step function. The thermal density matrix
is then translationally invariant:
\be
\rho_{\rm per}(\vec y,\vec y;\beta)
= \frac \beta2 \int\limits_0^\infty \frac{ds}{s}~\e^{-\frac 12 M^2 s}\,
\frac{1}{(2\pi s)^{d/2-1 }} \frac{\E}{4\pi \sinh {\frac{\E s}{2}}}\,
\Theta_3 \left(\nu\left \vert
\frac{i\beta^2 \E}{4\pi \tanh \frac{\E s}2}\right.\right)\,.
\label{finalrhoperiodic}
\ee
The choice of periodic vector potential~\rf{constantperiodic} differs from what
is considered above for a constant electric field since
\be
\dot{\vec  A}_{\rm per}(t)= \vec \E - \vec \E \beta\,\delta(t-\beta).
\ee

The lesson to be learned from this example is that the thermal partition
function in a constant electric field is trivial since the integration of
\eq{finalrho} over $\vec y$ picks up only the term of winding number $n=0$ and
the temperature dependence disappears. There are no excited states in the
external field and only the ground state exists as a stable configuration of
the charged particle. This feature occurs because there is no globally defined
periodic vector potential $\vec A(x^0)$ in this case as is required by the
standard formulation. Only in time-dependent backgrounds, such as
\eq{constantperiodic} which involves a time-localized point source of electric
field, does there exist excited configurations of the system. In this latter
case, the imaginary part of \eq{finalrhoperiodic} at the poles of the contour
integration, i.e. at $Fs=2\pi i\ell$ (integer $\ell$), yields the standard
(zero temperature) Schwinger probability amplitude for the creation of charged
particle pairs in scalar quantum electrodynamics. Note that there is a big
difference between constant electric and magnetic fields, because in the latter
case the field strength tensor does not have a component in the compactified
direction and the magnetic field would enter only in the pre-exponential
factors of the thermal density matrix, while the argument of the theta function
would be the same as without the field.

\subsection{Strings}

Consider an open superstring with independent $U(1)$ charges $e_1$ and $e_2$ at
its endpoints in an electromagnetic background. Here, strictly speaking, by an
Abelian background field we mean a field in an Abelian subgroup of the
Chan--Paton gauge group $O(32)$ of type-I superstring theory. We are interested
in the case $e_1\neq e_2$. However, unitarity requires the existence of neutral
strings in the spectrum. Consider a string scattering amplitude with the given
charges at the endpoints. An amplitude with an even number of external legs can
be sliced in many different ways into intermediate states. Some of these
intermediate states will consist of open strings with the charge $e_1$ or $e_2$
at both of its endpoints. An amplitude with an odd number of external legs
necessarily involves at least one neutral string in the scattering process.
Although these neutral string states will not be relevant to Debye screening
or the corresponding
T-dual D-brane dynamics, they do contribute to the total scattering
amplitudes which will involve sums of the form
\be
{\cal F}=\frac12\sum_{e_1,e_2\in{\cal Q}}{\cal F}(e_1,e_2)
\label{Ftotsum}\ee
where $\cal Q$ is the set of charges in the decomposition of the fundamental
representation of the open superstring Chan--Paton gauge group under the
embedding of $U(1)$ induced by the background electromagnetic field.

The open superstring partition function in a constant magnetic field at finite
temperature has been calculated in \cite{Tse98}. We will now show that charged
superstrings at finite temperature forbid constant electric fields. The bosonic
sector of this system can be described in first quantization by the
Polyakov path integral
\be
{\cal F}_b(e_1,e_2)=\int Dg_{ab}~\int Dx^\mu~\e^{-S_b[g_{ab},x^\mu]}
\label{polpath}\ee
where the action in the conformal gauge and in Euclidean spacetime is
\be
S_b=\frac1{4\pi\alpha'}\int\limits_\Sigma d^2z~\partial_zx_\mu\,\partial_{\bar
z}x^\mu+ie_1\oint\limits_{\partial_1\Sigma}A_\mu(x)~dx^\mu+ie_2\oint
\limits_{\partial_2\Sigma}A_\mu(x)~dx^\mu
\label{confaction}\ee
Finite temperature only affects the system when the string worldsheet $\Sigma$
can wrap the compact time direction, so the leading string diagram of interest
is the annulus\footnote{The M\"obius strip has only a single connected
boundary, so that only neutral strings contribute to the M\"obius amplitude. As
mentioned above, neutral superstrings will not play a significant role in the
subsequent analysis.} with local coordinates $z=\rho\e^{2\pi i\sigma}$ with
$0\leq\sigma\leq1$ and $a\leq\rho\leq1$, where $a=\e^{-t}$ is the Teichm\"uller
parameter of the annulus. We enforce the periodicity constraint in the
Euclidean time by substituting
\be
x^0(\rho,\sigma)~\mapsto~x^0(\rho,\sigma)+n\beta\sigma
\label{perconstr}\ee
and then later on summing the path integral over all winding numbers $n\in\bf
Z$.

We shall choose the periodic gauge field configuration \rf{gauge0}. With these
choices the action \rf{confaction} becomes
\bea
S_b&=&\frac1{4\pi\alpha'}\int d^2z~\partial_zx_\mu\,\partial_{\bar z}x^\mu+2\pi
i\nu n(e_2-e_1)+\frac{n^2\beta^2t}{8\pi^2\alpha'}\nonumber\\&
&+\,ie_2n\beta\int\limits_0^1d\sigma~\vec\E\cdot\vec
x(1,\sigma)-ie_1n\beta\int\limits_0^1d\sigma~\vec\E\cdot\vec
x(a,\sigma)\nonumber\\& &+\,ie_2\int\limits_0^1d\sigma~\vec\E\cdot\vec
x(1,\sigma)\,\partial_\sigma
x^0(1,\sigma)-ie_1\int\limits_0^1d\sigma~\vec\E\cdot\vec
x(a,\sigma)\,\partial_\sigma x^0(a,\sigma).
\label{annulusaction}\eea
There is a zero mode on the annulus given by $x^\mu=y^\mu={\rm const}$.
Normally, the action does not depend on it and integrating it out in the path
integral produces a factor $\beta V_{d-1}$, with $V_{d-1}$ the volume of space.
In the present case it contributes to \eq{polpath} the factor
\be
\beta\int d^{d-1}y~\e^{i(e_2-e_1)n\beta\vec \E\cdot\vec
y}=\beta\,(2\pi)^{d-1}\,\delta^{(d-1)}\Bigl((e_2-e_1)n\beta\vec \E\Bigr).
\label{0modeint}\ee
So unless $n=0$ (the zero temperature condition), or $\beta=0$ (the dual of the
zero temperature condition), or $e_1=e_2$ (neutral strings), the free energy of
the string gas is zero. Thus for charged strings at finite temperature in a
constant background electric field, the partition function picks up only the
$\beta$ independent $n=0$ sector.

This simple calculation exemplifies the fact that the translation
non-invariance of the finite temperature string theory produces a non-trivial
zero mode integration which localizes the string gas onto its ground state
configuration, unless the strings themselves are neutral. This is of course
anticipated from the example of the relativistic particle above and the fact
that the one-loop string free energy comes from an infinite tower of particle
states. In the following we will argue that this ground state localization is
evidence for Debye screening of electric fields and the appearance of a Debye
mass. We will then use T-duality to translate this into a statement about
D-brane dynamics.

\newsection{Debye Screening in Superstring Theory}

\subsection{Boundary State Formalism}

We will now compute the free energy of charged open superstrings in a constant
electric field by performing a modular transformation $t=1/s$ of the annulus
amplitude, which interchanges the roles of $\tau=-\ln\rho$ and $\sigma$, and
working in the closed string channel with a cylindrical worldsheet, i.e. the
boundary state formalism. In this representation, $0\leq\tau\leq s$
parametrizes the length of the cylinder so that the boundaries are at
$\tau=0,s$, and $0\leq\sigma\leq1$ parametrizes the closed string which
propagates through the cylinder with time coordinate $\tau$. The standard field
theoretic proper time is then $S=2\pi\alpha's$ (the same as for a particle).
Boundary states are coherent closed string states which insert a boundary on
the worldsheet and
enforce on it the appropriate boundary conditions. They are constructed by
applying the Boltzmann weight operator $\e^{-S_b}|_{\tau=0}$, constructed from
the action \rf{confaction}, to the vacuum state of the closed string Hilbert
space \cite{clny}. Finite temperature only affects the bosonic zero modes, and
we therefore consider only these contributions in detail. The bosonic part of
the boundary state which corresponds to one end of an open superstring is
created in part by the Wilson loop operator
\be
\Phi[x^\mu(\tau,\sigma)]=\exp
ie\left(\int\limits_0^1d\sigma~A_\mu(x)\,\partial_\sigma x^\mu-\chi(x)\right)
\label{bdryaction}\ee
where $e$ is the charge of the open string endpoint and $\chi(x)$ is the
appropriate Wu--Yang term described in Appendix A. For a constant electric
field $\vec\E$, we introduce the Euclidean angle $\cal E$ defined by
\be
\cos{\cal E}=\frac1{\sqrt{1+(2\pi\alpha'e\vec\E)^2}}
\label{calEdef}\ee
We shall choose coordinates in which $\vec\E=(\E,0,\dots,0)$ and the gauge
\rf{gauge1} with $c=1$. Then the Wu--Yang term is
\be
\chi(x)=\E x^1(\sigma=0)\left[x^0(\sigma=1)-x^0(\sigma=0)\right]
\label{wuyangcalE}\ee

To see what effects finite temperature imposes on the boundary states, we
consider \rf{bdryaction} as an operator on the closed string Hilbert space. The
closed string mode expansions are
\bea
x^0(\tau,\sigma)&=&y^0+\frac{2\pi i\alpha'n^0\tau}{s\beta}+
\frac{w^0\beta\sigma}{2\pi}+\sqrt{\alpha'}\sum_{n\neq0}
\frac1{in}\left(a_n^0\e^{-2\pi n(\tau/s+i\sigma)}+\tilde a_n^0\e^{-2\pi
n(\tau/s-i\sigma)}\right)\label{closedmodeexp0}\nonumber\\\vec x(\tau,\sigma)
&=&\vec y+\frac{i\alpha'\vec p\,\tau}s+\sqrt{\alpha'}\sum_{n\neq0}\frac1{in}
\left(\vec a_n\e^{-2\pi n(\tau/s+i\sigma)}+\vec{\tilde a}_n
\e^{-2\pi n(\tau/s-i\sigma)}\right)
\label{closedmodeexpj}\eea
in the sector of Kaluza--Klein momentum $n^0$ and winding number $w^0$ around
the compact Euclidean temperature direction. The Wu--Yang term
(\ref{wuyangcalE}) may then be written as
\be
\chi(x)=\frac{\E w^0\beta}{2\pi}\left[y^1+\frac{i\alpha'p^1\tau}s
+\sqrt{\alpha'}\sum_{n\neq0}\frac{\e^{-2\pi
n\tau/s}}{in}\left(a_n^1+\tilde a_n^1\right)\right]
\label{wuyangclosed}\ee
The total zero mode contribution to the action at $\tau=0$ is
\be
-iS_{w^0}=-iew^0\left(2\pi\nu-\frac{\beta\E}{2\pi}\,y^1\right)
\label{0modecontr}\ee
and it is straightforward to see that the oscillator part of
(\ref{wuyangclosed}) cancels the extra boundary integration in
(\ref{bdryaction}) which comes from the winding number term in the mode
expansion of $x^0$ in Eq. (\ref{closedmodeexp0}). The total oscillator
contribution to the action is therefore the standard one for open strings in
constant background electric fields. This simple calculation illustrates the
importance of incorporating the Wu--Yang term in compactifications involving
external fields.

In the bosonic sector, the boundary state $|B,e\rangle$ is required to satisfy
the rotated boundary conditions which follow from varying the worldsheet action
\rf{confaction} \cite{clny,BP92},
\bea
\left(\partial_\tau x^0+2\pi i\alpha'e\E\,\partial_\sigma
x^1\right)\Bigm|_{\tau=0}|B,e\rangle&=&0\nonumber\\\left(\partial_\tau
x^1-2\pi i\alpha'e\E\,\partial_\sigma
x^0\right)\Bigm|_{\tau=0}|B,e\rangle&=&0\nonumber\\\partial_\tau
x^j\Bigm|_{\tau=0}|B,e\rangle&=&0~~~~\forall j>1
\label{bdrycondns}\eea
This gives
\be
|B,e\rangle=\frac1{\cos{\cal E}}~|B_x,e\rangle^{(0)}\exp{\cal
O}({\cal E})|0\rangle_a|0\rangle_{\tilde a}~|B_{\rm gh}\rangle~|B_\psi,{\cal
E}\rangle
\label{bdrydef}\ee
where the bosonic zero-mode contributions to the boundary state are
\bea
|B_x,e\rangle^{(0)}&=&\sum_{w^0=-\infty}^\infty\e^{iS_{w^0}}\,
\left|n^0=0,w^0\right\rangle\prod_{j=1}^{d-1}
\left|k^j=0\right\rangle\nonumber\\&=&\sum_{w^0=-\infty}^\infty\e^{2\pi ie\nu
w^0}\,\left|n^0=0,w^0\right\rangle\left|k^1=-\frac{e\beta\E w^0}{2\pi}
\right\rangle\prod_{j=2}^{d-1}\left|k^j=0\right\rangle
\label{bdry0mode}\eea
Here $|n^0,w^0\rangle$ is the bosonic vacuum state which is normalized as
\be
\langle n',w'|n,w\rangle=\Phi_\beta\,\delta_{nn'}\delta_{ww'}
\label{vacnorm}\ee
where $\Phi_\beta$ is an appropriate volume factor which can be taken to be
the ``self-dual volume'' \cite{grv} of the compact direction that is fixed by
T-duality invariance to have the asymptotic behaviours
\be
\Phi_\beta\simeq\beta~~~~{\rm
for}~~\beta\to\infty~~,~~\Phi_\beta\simeq\frac{4\pi^2\alpha'}\beta~~~~{\rm
for}~~\beta\to0
\label{Phibetaasympt}\ee
The electric field dependent normalization in \eq{bdrydef} is the Born--Infeld
Lagrangian for the boundary gauge fields \cite{clny}, and $|B_{\rm gh}\rangle$
denotes the boundary state for the ghost and superghost degrees of freedom
which is unaffected by temperature and the electric field. The
$|k^j\rangle$ are the usual continuum momentum eigenstates in the $d-1$
uncompactified directions, while $|0\rangle_a$ and $|0\rangle_{\tilde a}$ are
Fock vacua for the bosonic closed string oscillator modes. The Bogoliubov
transformation
\be
{\cal O}({\cal
E})=\sum_{n=1}^\infty\left[\sum_{j=2}^{d-1}\frac1n\,a_{-n}^j\tilde
a_{-n}^j+\left(a_{-n}^0~,~a_{-n}^1\right)\pmatrix{\cos2{\cal E}&\sin2{\cal
E}\cr-\sin2{\cal E}&\cos2{\cal E}\cr}\pmatrix{\tilde a_{-n}^0\cr\tilde
a_{-n}^1\cr}\right]
\label{OcalE}\ee
encodes the boundary conditions (\ref{bdrycondns}) on the bosonic oscillatory
modes (see \cite{GG96,BCD,calkleb}, for example). It is unaffected by
temperature and so contributes the same quantity as at $T=0$. The fermionic
boundary coupling in the worldsheet superstring action is of the form
$ie\int_0^1d\sigma\,\overline\psi^\mu F_{\mu\nu}\psi^\nu$. The fermionic
boundary state $|B_\psi,{\cal E}\rangle$ is therefore also unaffected by finite
temperature, and it is given by a similar rotation as in \eq{OcalE} on the
fermionic oscillators and on the Ramond zero-mode states (see
\cite{clny,BCD,dfpslr} for the explicit expressions). However, finite
temperature breaks supersymmetry and modifies the GSO projection operator due
to the winding numbers $w^0$ around the compact temporal direction
\cite{attwitt}. This means that the sum over worldsheet spin structures
contains an extra weighting $(-1)^{w^0}$ for the $(-,+)$ spin structure in the
Neveu-Schwarz sector.

We shall now evaluate the one-loop superstring vacuum amplitude which is given
by the propagator
\bea
{\cal F}(e_1,e_2)&=&\frac1{\pi(2\alpha')^{d/2}}\,\left\langle B,e_2
\left|\frac1{L_0+\tilde
L_0-\eta}\right|B,e_1\right\rangle\nonumber
\\&=&\frac1{\pi^2(2\alpha')^{d/2}}\,
\int\limits_0^\infty ds~\left\langle
B,e_2\Bigm|\e^{-\pi s(L_0+\tilde L_0-\eta)}\Bigm|B,e_1\right\rangle
\label{bosprop}\eea
where $s$ is the modular parameter of the cylinder and the normal ordering
intercept is $\eta=2$ in the bosonic sector, while $\eta=\frac12$ in the
fermionic Neveu-Schwarz sector and $\eta=0$ in the Ramond sector. The bosonic
part of the closed string Hamiltonian is
\be
L_0^b+\tilde
L_0^b-2=\frac{8\pi^2\alpha'(n^0)^2}{\beta^2}+
\frac{(w^0)^2\beta^2}{2\pi^2\alpha'}+
2\alpha'\vec p^{\,2}+N_b+\tilde N_b-2
\label{closedstrham}\ee
where $N_b$ and $\tilde N_b$ are the usual number operators for the bosonic
oscillatory and ghost modes. After some algebra we find (setting $d=10$)
\bea
{\cal F}(e_1,e_2)&=&\frac{16\sin\pi\epsilon}
{(2\pi\alpha')^5\cos{\cal E}_1\cos{\cal E}_2}\,\Phi_\beta V_8
\sum_{w^0=-\infty}^\infty\e^{2\pi i\nu w^0(e_1-e_2)}\,\delta
\left((e_1-e_2)w^0\beta\E\right)\nonumber\\& &
\times\,\int\limits_0^\infty ds~\e^{-\frac s{2\pi\alpha'}(w^0)^2\beta^2
\left[1+(2\pi\alpha'e_1\E)^2\right]}\,\frac{\Theta_1'(0|is)^{-3}}
{\Theta_1(\epsilon|is)}\nonumber\\& &\times\left[\Theta_3(\epsilon|is)\,
\Theta_3(0|is)^3-(-1)^{w^0}\,\Theta_4(\epsilon|is)\,
\Theta_4(0|is)^3-\,\Theta_2(\epsilon|is)\,
\Theta_2(0|is)^3\right]\nonumber\\& &
\label{supervacampl}\eea
where $\Theta_\alpha$ denote the usual Jacobi theta functions which can be
expressed in terms of the triple product formulas
\bea
\Theta_1(\nu|i\tau)&=&2\e^{-\pi\tau/4}\,\sin\pi\nu\prod_{n=1}^\infty
\left(1-\e^{-2\pi n\tau}\right)\left(1-\e^{-2\pi n\tau+2\pi i\nu}\right)
\left(1-\e^{-2\pi n\tau-2\pi i\nu}\right)\nonumber\\
\Theta_2(\nu|i\tau)&=&2\e^{-\pi\tau/4}\,\cos\pi\nu\prod_{n
=1}^\infty\left(1-\e^{-2\pi n\tau}\right)\left(1+\e^{-2\pi n
\tau+2\pi i\nu}\right)\left(1+\e^{-2\pi n\tau-2\pi i\nu}\right)
\nonumber\\\Theta_3(\nu|i\tau)&=&\prod_{n=1}^\infty\left(1-\e^{-2\pi n\tau}
\right)\left(1+\e^{-(2n-1)\pi\tau+2\pi i\nu}\right)\left(1+\e^{-(2n-1)
\pi\tau-2\pi i\nu}\right)\nonumber\\\Theta_4(\nu|i\tau)&=&\prod_{n=1}^\infty
\left(1-\e^{-2\pi n\tau}\right)\left(1-\e^{-(2n-1)\pi\tau+2\pi i\nu}
\right)\left(1-\e^{-(2n-1)\pi\tau-2\pi i\nu}\right)
\label{Thetadefs}\eea
and $\Theta_1'(0|i\tau)=\partial_\nu\Theta_1(\nu|i\tau)|_{\nu=0}$. In
\eq{supervacampl} we have introduced the twist parameter
\be
\pi\epsilon={\cal E}_1-{\cal E}_2
\label{epsilondef}\ee
and used the normalization
\be
\delta^{(d-2)}(0)=\frac{V_{d-2}}{(2\pi)^{d-2}}
\label{delta0norm}\ee
A factor of $\Theta_1'(0|it)^{-4}$ in \eq{supervacampl} comes from the bosonic
oscillator modes in the eight directions transverse to the 0--1 plane, and a
factor $\Theta_1'(0|it)$ comes from the ghosts while the rotated light-like
degrees of freedom contribute $\Theta_1(\epsilon|it)$. These latter two
contributions cancel each other in the neutral string limit $e_1=e_2$ or in the
absence of
the electric field. The sum of theta functions in the last line of
\eq{supervacampl} similarly comes from the fermionic oscillators with the first
two representing the contributions from the Neveu-Schwarz spin structures and
the last one the Ramond sector component. The delta function comes from the
overlap $\langle k_2^1|k_1^1\rangle$ after using the orthogonality condition
(\ref{vacnorm}).

We see once again that the free energy of charged superstrings is independent
of temperature. The triviality in the present case arises because the operator
\rf{0modecontr} translates the momentum $k^1$ in the boundary states 
\rf{bdry0mode} by the {\it dual} momentum in the temporal direction.
In the case of neutral superstrings, whereby $e_1=e_2=e$
and $\epsilon=0$, we may use the Jacobi abstruse identity
\be
\Theta_3(0|is)^4-\Theta_4(0|is)^4=\Theta_2(0|is)^4
\label{abstruse}\ee
and the series expansion
\be
\Theta_2(\nu|i\tau)=\sum_{q=-\infty}^\infty\e^{-\pi(2q+1)^2\tau/4
+i\pi(2q+1)\nu}
\label{Theta2series}\ee
to write the amplitude (\ref{supervacampl}) in the form
\bea
{\cal F}(e,e)&=&\frac{32}{(2\pi\alpha')^5}\,\Phi_\beta
V_9\left[1+(2\pi\alpha'e\E)^2\right]\nonumber\\& &\times\,\int\limits_0^\infty
ds~\Theta_2\left(0\left|\frac{i\beta^2s}{2\pi^2\alpha'}\,
\left[1+(2\pi\alpha'e\E)^2
\right]\right.\right)\,\left[\frac{\Theta_4(0|is)}{\Theta_1'(0|is)}\right]^4
\label{neutralampl}\eea
\eq{neutralampl} coincides with the standard expression for the partition
function of the neutral superstring in a constant background electric field
\cite{FFdI,Tse98} (after the modular transformation $s=1/t$ and Wick rotation
to the Minkowski electric field \rf{FvsE}). Taking into account of the fact
that for a given set of charges $e_1,e_2$ the spectrum must contain neutral
strings (c.f. \eq{Ftotsum}), we arrive at the total one-loop annulus amplitude
in the closed string representation,\footnote{To this expression should be
added the contribution of neutral string states from the M\"obius amplitude.
The total result would have the same qualitative properties.}~\bea
{\cal F}_{\{e_1,e_2\}}&=&\frac{16}{(2\pi\alpha')^5}\,
\Phi_\beta V_9\,\int\limits_0^\infty
ds~\frac1{\Theta_1'(0|is)^4}\nonumber\\&
&\times\,\left\{\Theta_4(0|is)^4\sum_{a=1,2}\left[1+(2\pi\alpha'e_aF)^2
\right]\,\Theta_2\left(0\left|\frac{i\beta^2s}{2\pi^2\alpha'}
\left[1+(2\pi\alpha'e_aF)^2\right]\right.\right)\right..\nonumber\\& &
\left.+\,2\sin\pi\epsilon\,\sqrt{1+(2\pi\alpha'e_1F)^2}\,\sqrt
{1+(2\pi\alpha'e_2F)^2}~\Theta_1\left(\left.\frac\epsilon2\right|is
\right)^4\,\frac{\Theta_1'(0|is)}{\Theta_1(\epsilon|is)}\right\}
\nonumber\\& &
\label{totannampl}\eea
where we have used the formula
\be
\Theta_3\left(\nu\left\vert{i\tau}\right.\right)
\,\Theta_3\left(0\left\vert{i\tau}\right.\right)^3
-\Theta_4\left(\nu\left\vert i\tau\right.\right)
\,\Theta_4\left(0\left\vert{i\tau}\right.\right)^3
-\Theta_2\left(\nu\left\vert{i\tau}\right.\right)
\,\Theta_2\left(0\left\vert{i\tau}\right.\right)^3
=2\,\Theta_1\left(\left.\frac \nu2 \right\vert i\tau\right)^4
\ee
which is a consequence of the Riemann identity \cite{GG96}. Upon examining the
region of convergence of the $s\to\infty$ modular parameter integration in
\eq{totannampl}, we arrive at the usual modification \cite{FFdI,Tse98} of the
critical Hagedorn temperature of the free open superstring gas due to the
presence of the electric field,
\be
T_{\rm H}(F)=\frac1{2\pi k_{\rm B}\sqrt{2\alpha'}}\,
\sqrt{1+(2\pi\alpha'e_0F)^2}
\label{Hagtemp}\ee
where $e_0$ is the minimum unit of electric charge.

\subsection{Effective Action for the Polyakov Loop}

One can readily see from \eq{supervacampl} the correlation between the
dependence of the correlator of Polyakov loops on the temporal gauge field
component $A_0$ and the vanishing of the electric field. For a charged string
at finite temperature, in order for the first exponential in \eq{supervacampl}
to depend on $A_0$, the delta function must force $\E=0$. Taking this zero
field limit, we can use \eq{supervacampl} to see the occurrence of Debye
screening from the fact that there is a non-trivial effective action for the
Polyakov loop operator. Instead of \eq{bdryaction}, we will consider a more
general condensate of photon
vertices defined by the path ordered Polyakov loop operator
\be
{\cal P}[A]=\prod_\alpha\,{\rm Tr~P}\,\exp
i\oint\limits_{\partial_\alpha\Sigma}A_\mu(x)~dx^\mu
\label{polloopdef}\ee
associated with the open superstring Chan--Paton gauge group. The boundary of
the string worldsheet is in general a set of disconnected closed loops,
$\partial\Sigma=\bigcup_\alpha\partial_\alpha\Sigma$. The screening of electric
fields owes to special properties of the finite temperature theory. As all
observables must be periodic in $x^0$, at least up to a gauge transformation,
the temporal gauge field $A_0(x)$ has a special status, since not all gauge
transformations are allowed. The gauge transformations $U$ which are allowed
are those for which $U$ is periodic up to an element of the center of the gauge
group,
\begin{equation}
U(x^0+\beta,\vec x)=\e^{i\theta}\,U(x^0,\vec x)
\end{equation}
This symmetry is a result of the fact that all string states transform in
either the adjoint or other zero $N$-ality representations of the Chan--Paton
gauge group.\footnote{Note that $\theta=\pi$ for type-I superstrings, but later
on we shall consider analogous statements for D-branes in which the
gauge group will be $U(N)$ for some $N$ and $\theta$ is arbitrary.} Related to
this symmetry is the fact that screening occurs only for fields in the $su(N)$
subalgebra of $u(N)$ where $\sum_b(A_0)_{bb}=0~{\rm mod}\,N$. The true gauge
group is not $U(N)$ but rather the quotient of $U(N)$ by its center, or
$U(N)/U(1)=SU(N)/{\bf Z}_N$.

This feature results in two facts. First of all, we can always choose a gauge
where $A_0$ is time-independent and diagonal. However, at finite temperature
one cannot completely remove the $A_0$'s by the residual Abelian gauge
invariance because of the existence of the non-trivial gauge invariant holonomy
\rf{polloopdef}. Secondly, there is a global symmetry under the simultaneous
translation of all diagonal elements of $A_0$,
\begin{equation}
A_0^{aa}(\vec x)\rightarrow A_0^{aa}(\vec x)+\theta
\end{equation}
This global symmetry can be restricted to the ${\bf Z}_N$ subgroup of $U(1)$
where $\theta=2\pi n/\beta$ (integer $n$). It is a result of the existence of
large gauge transformations,
\be
A_\mu\rightarrow U\left( A_\mu -i\partial_\mu\right)U^{-1}
\ee
where $U=\e^{2\pi inx^0E^b/\beta}$ and $E^b$ is the matrix whose only
non-vanishing entry is $(E^b)_{bb}=1$. This gauge transform is periodic in
$x^0$ and under which
\be
A_0\rightarrow A_0+\frac{2\pi n}\beta\,E^b
\ee
The effective potential should exhibit this invariance. In a static diagonal
gauge where the worldsheet boundary wraps the periodic time direction, the
Polyakov loop operator is given by ${\cal P}=\sum_b\e^{i\beta a^b(x)}$. Then
$F(x)\equiv -\beta^{-1}\ln\langle{\cal P}(x)\rangle$ is the free energy that
would be required to introduce a heavy fundamental representation quark into
the system and thereby gives information about confinement.

For example, via a gauge transformation we can consider the background field
\begin{equation}
A_0^{bc}(x)=\delta^{bc}\,a^b_0
\end{equation}
where $a_0^b$ are constants. When the worldsheet $\Sigma$ is an annulus, the
effective action is a sum over the differences between all Abelian $U(1)^N$
charges $a_0^b$ located at the two boundaries of $\Sigma$. This is easily
incorporated into the boundary state formalism described above, and from
\eq{supervacampl} we find that the effective potential for the
Polyakov loop operator is given by
\be
\Gamma[A]=\frac{32}{(2\pi\alpha')^5}\,\Phi_\beta
V_9\int\limits_0^\infty\frac{dt}{t^6}~\sum_{b,c}\Theta_2\left(\frac\beta\pi
\left(a_0^b-a_0^c\right)\left|\frac{i\beta^2}{2\pi^2\alpha't}\right.\right)
\left[\frac{\Theta_2(0|it)}{\Theta_1'(0|it)}\right]^4
\label{effactionpol}\ee
where we have made a modular transformation $s=1/t$ and used the Poisson
resummation formula
\be
\frac1\tau\,\frac{\Theta_4(\nu|i\tau)}{\Theta_1'(0|i\tau)}=\e^{-\pi
\nu^2/\tau}\,\frac{\Theta_2\left(\frac{i\nu}\tau\left|\frac
i\tau\right.\right)}{\Theta_1'\left(0\left|\frac i\tau\right.\right)}
\label{PoissonT24}\ee
The case of a single charged superstring is associated with gauge group $U(2)$
and $a_0^b=2\pi\nu e_b$ in \eq{effactionpol}.

The effective action can be written more explicitly as a sum over superstring
states and temporal winding numbers. The integration over the Teichm\"uller
parameter $t$ of the annulus may be evaluated by expanding the ratio of theta
functions in \eq{effactionpol} using the formula
\be
8\prod_{n=1}^\infty\left(\frac{1+\e^{-2\pi nt}}{1-\e^{-2\pi
nt}}\right)^8=\sum_{N=0}^\infty d_N\,\e^{-2\pi Nt}
\label{statesum}\ee
where $d_N$ is the degeneracy of superstring states at level $N$. Using
\eq{Theta2series} this gives
\be
\Gamma[A]=2^6(\alpha')^{-5/2}\,\pi^2\,\frac{\Phi_\beta
V_9}{(\pi\beta)^5}\,\sum_{N=0}^\infty
d_N\,N^{5/2}\sum_{q=-\infty}^\infty\frac{K_5\left(\beta|2q+1|\sqrt{\frac
N{\alpha'}}\right)}{|2q+1|^5}\,\sum_{b,c}\e^{i\beta(2q+1)(a_0^b-a_0^c)}
\label{effactionexpl}\ee
where $K_5(z)$ is the irregular modified Bessel function of order 5. Using the
asymptotic behaviours
\be
K_5(z)\simeq\frac{3\cdot2^7}{z^5}~~~~{\rm for}~~|z|\to0~~,~~
K_5(z)\simeq\e^{-z}\,\sqrt{\frac\pi{2z}}~~~~{\rm for}~~|z|\to\infty
\label{K5asympt}\ee
the low temperature limit of \eq{effactionexpl} picks up the
lowest $|2q+1|=1$ winding modes giving
\be
\Gamma[A]\simeq2^6\,\Phi_\beta V_9
\left(\frac{3\cdot2^{10}}{\pi^4\beta^{10}}+\frac{\sqrt2\pi
d_1}{(\alpha')^{9/4}}
\,\frac{\e^{-\beta/\sqrt{\alpha'}}}{(\pi\beta)^{11/2}}\right)\sum_{b,c}
\cos\left(\beta(a_0^b-a_0^c)\right)~~~~~~{\rm for}~~\beta\to\infty
\label{effactionlowT}\ee
The first term in \eq{effactionlowT} comes from the lowest lying $N=0$ states
which have degeneracy $d_0=8$, while the second term comes from the first
excited $N=1$ levels.

\subsection{Some Properties of the Effective Action}

The coefficient of the expansion of \eq{effactionexpl} or \eq{effactionlowT} to
order $A_0^2$ yields an expression for the Debye screening mass $\mu^2/2$. The
leading contribution is given in \eq{debyemass} and it is the same as the Debye
mass that one would calculate in ordinary ten-dimensional Yang-Mills theory.
This is due to the fact that the dominant term in \eq{effactionlowT} at low
temperatures comes from only the particle-like excitations of the strings. The
leading stringy corrections are exponentially suppressed by factors
$\e^{-\beta/\sqrt{\alpha'}}$. In the low temperature regime these corrections
play no role in the Debye screening. This exponential suppression is a result
of supersymmetry which leads to cancellation of the leading order stringy
corrections to the mass $\mu$. It is interesting that stringy effects only play
a role at temperatures near the Hagedorn transition. Notice also that
$\mu^2\propto T/T_{\rm H}(0)$, so that well below the critical temperature the
Debye mass is small and the electric fields become more and more long-ranged.
This suggests that there should exist gauge field configurations with a
relatively ``mild'' time dependence for which there is no screening of the
(time dependent) electric fields (such as the example given at the end of
section 2.1).

The functional $\Gamma[A]$ above is the leading term in the derivative
expansion of the full effective potential for the gauge field
$a(x)^b=e_b\,a(x)$. As discussed in section 1, it should be added to the
tree-level Born--Infeld Lagrangian for the field $A_0(\vec x)$, thereby
determining the one-loop, temperature corrections to the gauge field effective
action of the superstring theory. The solutions of the resulting equations of
motion then determine the allowed, on-shell gauge field configurations which
lead to a conformally-invariant theory at finite temperature. At low
temperatures, the modified Born--Infeld action is given by the free energy
\be
F[a]=\frac{\sum_be_b^2}{(2\pi)^5(\alpha')^3g_s}
\,\int d^{9}x~\left\{\sum_{b}\frac{\sqrt{1+\left(
2\pi \alpha'e_b\vec\nabla a(x)\right)^2}}{(2\pi\alpha')^2\sum_ce_c^2}+
\sum_{b,c}\frac{\mu^2}{\beta^2}\,\cos\Bigl(\beta(e_b-e_c)a(x)\Bigr)
+\ldots\right\}
\label{cgas}\ee
We can think of the modified Born--Infeld action (\ref{cgas}) as a
generalization of the sine-Gordon theory representation of the classical
Coulomb gas where the standard kinetic term $(\vec\nabla a)^2$ is replaced by
the Born--Infeld Lagrangian. The sine-Gordon theory has well-known soliton
solutions, which correspond to solitary waves of the plasma phase of the
Coulomb gas. It is interesting to note that the Born--Infeld generalization of
the Coulomb gas (\ref{cgas}) also has solitons. Consider the ansatz where
$a(x)$ depends on only one space coordinate. Then the non-linear equation for
the extrema of (\ref{cgas}) is solved by the function $a(x)$ which is obtained
from the integral
\begin{equation}
\int\frac{\pi\beta(2\pi\alpha')^{-1/2}~da}{2\mu\Bigl(\sum_be_b^2\Bigr)
\left(\sum_{b,c}\left|\sin\Bigl((e_b-e_c)\beta a/2\Bigr)
\right|\right)\sqrt{\frac{\mu^2g_s\sum_be_b^2}{(2\pi)^5(\alpha')^3\beta^2}
\,\sum_{b,c}\sin^2\Bigl((e_b-e_c)\beta a/2\Bigr)+1}} = x
\label{soliton}
\end{equation}
These solitons also exist in gauge field theories as ${\bf Z}_N$ domain walls.

\newsection{T-duality and Moving D-branes}

The boundary state formalism of the previous section presents the most
efficient way to map the results for the electric field problem into statements
about D-brane dynamics at finite temperature. String theory in background gauge
fields is related by T-duality to open string theory with moving D-branes.
The prerequisite for this relation is translation invariance of the external
fields in the directions which will be compactified. This means that the gauge
fields should be translationally invariant up to a gauge transformation.
Generally this is hard to do in first quantization because the
required Wu--Yang terms spoil the translation invariance. But, as we now
demonstrate, the boundary state formalism gives a precise prescription for
obtaining the T-dual D-brane picture.

For this, we compactify the Neumann string coordinates $x^i\equiv x_{\rm N}^i$
for $i=1,\dots,d-p-1$ on circles of circumferences $L_i$. This modifies the
closed string mode expansions \rf{closedmodeexpj} along the first $d-p-1$
spatial directions to
\be
x_{\rm N}^i(\tau,\sigma)=y^i+\frac{2\pi i\alpha'n^i\tau}{sL_i}
+\frac{w^iL_i\sigma}{2\pi}+
\sqrt{\alpha'}\sum_{n\neq0}\frac1{in}\left(a_n^i\e^{-2\pi n(\tau/s+i\sigma)}
+\tilde a^i_n\e^{-2\pi n(\tau/s-i\sigma)}\right)
\label{x1modeexpL}\ee
where $n^1$ is the Kaluza--Klein index and $w^1$ the winding number around the
compactified direction along which the electric field lies. The Wu--Yang term
in this case is given in Appendix A and, for the gauge choice of section 3.1,
it coincides with \eq{wuyangclosed}. The zero mode contribution to the Wilson
line integral in \eq{bdryaction} at $\tau=0$ is modified to
\be
-i\tilde S_{w^0,w^1}=-2\pi ie\nu w^0-\frac{ieF}{2\pi}\left(
y^0L_1w^1-y^1\beta w^0\right)
\label{zeromodecomp}\ee
which now contains the rotation generator on the spacetime torus (again
parametrized by {\it dual} momenta). The
oscillator contributions are again unchanged, and in the boundary states
\rf{bdrydef} the zero mode momentum states in \eq{bdry0mode} are now also
discretized in the first $d-p-1$ spatial directions. Upon applying the zero
mode operator $\e^{i\tilde S_{w^0,w^1}}$ to the Fock vacuum, which shifts the
total light-like momentum, we find that the only modification to the boundary
state is again in the bosonic zero mode part which now becomes
\bea
|B_x,e\rangle^{(0)}&=&\sum_{w^0,w^1=-\infty}^\infty\e^{2\pi ie\nu
w^0}\left|n^0=\frac{e\beta
L_1Fw^1}{2\pi}~,~w^0\right\rangle\left|n^1=-\frac{e\beta
L_1Fw^0}{2\pi}~,~w^1\right\rangle\nonumber\\&
&\times\,\prod_{j=2}^{d-p-1}~\sum_{w^j=-\infty}^\infty|n^j=0,w^j\rangle~
\prod_{i\geq d-p}|k^i=0\rangle
\label{bdry0modecomp}\eea
The shifts in the Kaluza--Klein integers in \eq{bdry0modecomp} imply that the
electric field must be quantized as
\be
F=\frac{2\pi}{e_0\beta L_1}\,N
\label{efieldquant}\ee
where $N$ is an integer. The constraint \rf{efieldquant} is derived in Appendix
A from a purely mathematical point of view and is shown to be a topological
quantization condition associated with the presence of the Wu--Yang term.

Now we apply a T-duality transformation along the compact spatial directions
and write down the corresponding boundary state for a moving D$p$-brane. This
mapping interchanges the Neumann and Dirichlet boundary conditions for the open
string along the first $d-p-1$ spatial directions. The new string coordinates
$x_{\rm D}^j$ take values on the dual circles of circumferences
\be
\tilde L_j=\frac{4\pi^2\alpha'}{L_j}
\label{dualcircum}\ee
and they are defined by the equations
\be
\partial_\tau x_{\rm D}^j=-i\,\partial_\sigma x_{\rm
N}^j~~~~,~~~~\partial_\sigma x_{\rm D}^j=i\,\partial_\tau x_{\rm
N}^j~~~~~~j=1,\dots,d-p-1
\label{xDdef}\ee
The boundary state conditions \rf{bdrycondns} now become
\bea
\partial_\tau\left(x^0-vx_{\rm
D}^1\right)\Bigm|_{\tau=0}|Dp,y_0;v\rangle&=&0\nonumber\\\partial_\sigma
\left(x_{\rm D}^1+vx^0\right)\Bigm|_{\tau=0}|Dp,y_0;v\rangle&=&0\nonumber\\
\partial_\sigma x_{\rm D}^j\Bigm|_{\tau=0}|Dp,y_0;v\rangle&=&0~~~~
j=2,\dots,d-p-1\nonumber\\\partial_\tau x^
i\Bigm|_{\tau=0}|Dp,y_0;v\rangle&=&0~~~~i\geq d-p
\label{dualbdrycondns}\eea
where
\be
v=2\pi\alpha'eF=\frac{eN}{e_0}\,\frac{\tilde L_1}\beta
\label{velocity}\ee
is the velocity of the string endpoint in the direction 1 transverse to the
D-brane. The constraint \rf{velocity} can be thought of as
momentum quantization along the compact boost direction, with the quantum of
velocity equal to the speed in going once around the dual circle in a single
unit of Matsubara time. The Dirichlet boundary conditions in
\eq{dualbdrycondns} may be
alternatively expressed by setting the operators equal to fixed positions
$y_0^i$ at $\tau=0$ which we interpret as the transverse coordinates of the
D$p$-brane. Note that the angle $\cal E$ defined in \eq{calEdef} is interpreted
in the dual picture as the Euclidean rapidity of the D-brane motion, so that
the twist parameter \rf{epsilondef} becomes the relativistic composition of the
velocities of two branes \cite{Bac96}.

The boundary conditions \rf{dualbdrycondns} are solved by acting on Fock vacua
with the dual version of the boundary operator \rf{bdryaction} obtained by
substituting $x_{\rm N}^i\mapsto x_{\rm D}^i$ for $i=1,\dots,d-p-1$:
\be
\tilde\Phi[x^\mu(\tau,\sigma)]=\exp\left(2\pi i\tilde\nu
w^0-\frac1{2\pi\alpha'}\int\limits_0^1d\sigma~y_1(x^0)\,\partial_\tau x_{\rm
D}^1+\frac1{2\pi\alpha'}\,\tilde\chi(x)\right)
\label{dualbdryaction}\ee
where $\tilde\nu=e\nu$ is now interpreted as a gauge field living in the
D-brane worldvolume, and
\be
y_1(x^0)=vx^0
\label{branecoord}\ee
is the D-brane trajectory induced by the gauge potential transverse to the
brane, which produces in \eq{dualbdryaction} the standard boundary vertex
operator for a moving D-brane \cite{leigh,calkleb}.
The dual of the Wu--Yang term \rf{wuyangcalE} is given by
\be
\tilde\chi(x)=vx_{\rm D}^1(\sigma=0)\left[x^0(\sigma=1)-x^0(\sigma=0)\right]
\label{dualwuyang}\ee
which in the closed string parametrization at $\tau=0$ is obtained from
\eq{wuyangclosed} by reflecting the left-moving oscillators $\tilde
a_n^1\mapsto-\tilde a_n^1$. For the case of D-brane motion, the Wu--Yang term
\rf{dualwuyang} is associated with maintaining reparametrization invariance of
the D-brane worldvolume along the boosted direction at finite temperature. Its
effect when the operator \rf{dualbdryaction} is written in terms of closed
string mode expansions is identical to that of the electric field problem. Note
that off-shell the field \rf{dualwuyang} has a very complicated, non-local
form. It is this feature which makes a path integral calculation technically
problematic, in contrast to the boundary state formalism which uses on-shell
string embedding fields.

Upon decompactifying the dual circles in every spatial direction but the
boosted one, we arrive at the D$p$-brane boundary state
\cite{BCD,calkleb,dfpslr}
\be
|Dp,y_0;v\rangle=\frac{\sqrt\pi}2\left(2\pi\sqrt{\alpha'}\right)^{3-p}
\frac1{\cos{\cal E}}~|Dp_x,y_0;v\rangle^{(0)}\exp\tilde{\cal O}({\cal E})
|0\rangle_a|0\rangle_{\tilde a}~|B_{\rm gh}\rangle~|B_\psi,{\cal E}\rangle
\label{dualbdrydef}\ee
which represents the source for the closed string modes emitted by the brane.
The numerical normalization factor in \eq{dualbdrydef} is the D$p$-brane
tension \cite{fpslr}, while the factor of $\cos{\cal E}$ inserts the
appropriate Lorentz contraction factor. The operator $\tilde{\cal O}({\cal E})$
is obtained from \eq{OcalE} by changing the sign of $\tilde a_n^i$ for
$i=1,\dots,d-p-1$. The ghost and fermionic boundary states are identical to
those used in section 3.1, while the bosonic zero mode part of the boundary
state is now
\be
|Dp_x,y_0;v\rangle^{(0)}=\sum_{w^0,\tilde w^1=-\infty}^\infty\e^{i\tilde
S_{w^0,\tilde
n^1}}\,\prod_{j=1}^{d-p-1}\delta\left(y^j-y_0^j\right)
\,|n^0=0,w^0\rangle|\tilde
n^1=0,\tilde w^1\rangle\prod_{i=2}^{d-1}|k^i=0\rangle
\label{dual0modebdry}\ee
where $\tilde n^1=w^1$ is the dual momentum and $\tilde w^1=n^1$ the dual
winding number around the compact boost direction. As before the closed string
zero-mode operator \rf{zeromodecomp} accounts for the boosted boundary
conditions in the light-like plane, while the transverse Dirichlet boundary
conditions in \eq{dualbdrycondns} are enforced by the delta-function operators
in \eq{dual0modebdry} which are defined by the Fourier expansions
\bea
\delta\left(y^1-y_0^1\right)&=&\sum_{\tilde n^1=-\infty}^\infty\e^{2\pi i\tilde
n^1\left(y^1-y_0^1\right)/\tilde L_1}\nonumber\\
\delta\left(y^j-y_0^j\right)&=&\int\limits_{-\infty}^\infty\frac{dq_j}{2\pi}~
\e^{iq_j\left(y^j-y_0^j\right)}~~~~~~j=2,\dots,d-p-1
\label{deltaopdef}\eea
These delta-functions have the effect of introducing extra terms, proportional
to $|\vec y|^2$, into the energy levels coming from the masses of the open
string excitations which stretch between a pair of branes at separation
$|\vec y|$. The
compactness of the boost direction is required since the operator
\rf{dualbdryaction} shifts the temporal Kaluza--Klein momentum $n^0$ by
$vq^1/2\pi$, which is only an integer when the momentum $q^1$ is discretized
and the velocity is quantized according to \eq{velocity}. Likewise, the
operator $\e^{i\tilde S_{w^0,\tilde n^1}}$ shifts the Kaluza--Klein momentum
$\tilde n^1$ along the boost direction by $\tilde L_1vw^0/\tilde\beta$, which
with the topological quantization condition \rf{velocity} is an integer only
for $w^0=0$. The Dirichlet zero mode boundary states are thus
\bea
|Dp_x,y_0;v\rangle^{(0)}&=&\sum_{\tilde w^1=-\infty}^\infty~\sum_{\tilde
n^1=-\infty}^\infty\e^{-2\pi i\tilde n^1y_0^1/\tilde
L_1}\,\left|n^0=\frac{\beta v\tilde n^1}{\tilde
L_1}~,~w^0=0\right\rangle\left|\tilde n^1,\tilde w^1\right\rangle\nonumber\\&
&\times\,\prod_{j=2}^{d-p-1}\int\limits_{-\infty}^\infty\frac{dq_j}{2\pi}~
\e^{-iq_jy_0^j}\,|q_j\rangle~\prod_{i\geq d-p}|k^i=0\rangle
\label{Dbdry0mode}\eea

It is important to realize that the form \rf{Dbdry0mode} for the boundary state
is only valid at finite temperature and velocity. For a static D$p$-brane the
boost direction can be decompactified and the
boundary state is instead given from \eq{dual0modebdry} as
\be
|Dp_x,y_0;v=0\rangle^{(0)}=\sum_{w^0=-\infty}^\infty\e^{2\pi i\tilde\nu
w^0}\,|n^0=0,w^0\rangle\prod_{j=1}^{d-p-1}\int\limits_{-\infty}^\infty
\frac{dq_j}{2\pi}~\e^{-iq_jy_0^j}\,|q_j\rangle~\prod_{i\geq d-p}|k^i=0\rangle
\label{Dpstatev0}\ee
which produces a non-trivial winding mode dependence in the temperature
direction. If we had used instead of \eq{velocity} the dual quantization
condition $\tilde v=\tilde N\tilde\beta/\tilde L_1$, $\tilde N\in{\bf Z}$,
in \eq{dual0modebdry}, then the Kaluza--Klein momenta of the boundary states
would be $n^0=0$ and $\tilde n^1=-w^0\tilde v\tilde L_1/\tilde\beta$ while
their winding numbers would be arbitrary.
This state would then coincide with \eq{Dpstatev0} in the zero velocity limit.
However, as shown in Appendix A, \eq{velocity} has a deep mathematical origin
and is the correct one to use for the present problem. It arises from
the mathematical property that the light-like torus has non-trivial \v{C}ech
cohomology which makes it impossible to define a globally non-singular vector
potential on it with non-vanishing flux. We see then that the zero mode
boundary state \rf{Dbdry0mode} is the appropriate T-dual of \eq{bdry0mode}
which contains as well a sort of temperature duality transformation, in the
sense that the boosted boundary conditions forbid closed string windings around
the temporal direction but now the closed string energies are non-vanishing.

We see that from the onset the D$p$-brane boundary state represents
only the trivial $w^0=0$ state at finite temperature. This property may be
attributed to the fact that the closed string operator which is used to boost
a static D-brane boundary state \cite{BCD} doesn't commute with the 
compactification of the boost plane and thereby produces a non-trivial
projective phase. Indeed, the free energy
\be
\tilde{\cal F}(v_1,v_2)=\frac2{\pi(2\alpha')^{\frac d2+3-p}}\,
\left\langle Dp,y_0^{(2)};v_2\left|\frac1{L_0+\tilde
L_0-\eta}\right|Dp,y_0^{(1)};v_1\right\rangle
\label{freev}\ee
may be computed using \eq{Dbdry0mode} and the results of section 3.1. Here an
extra factor of 2 has been inserted to take into account of the fact that one
can interchange the roles of the two endpoints of an oriented string
\cite{Pol95}. After computing the discrete overlaps of states using the
orthogonality conditions, one can safely take the decompactification limit
$\tilde L_1\to\infty$ with $\tilde w^1=0$ and $q_1=2\pi\tilde n^1/\tilde L_1$ a
fixed continuum momentum variable. Integrating over the momenta $q_j$ for
$2\leq j\leq d-p-1$, we arrive at
\bea
\tilde{\cal
F}(v_1,v_2)&=&2^{p/2}\,(2\pi)^{4-2p}\,\left(4\pi^2\alpha'\right)^{-p/2}
\,\sqrt{1+v_1^2}\,\sqrt{1+v_2^2}
{}~\Phi_\beta V_p\nonumber\\&
&\times\,\int\limits_0^\infty\frac{ds}{s^{4-p/2}}~
\e^{-b^2/2\pi\alpha's}\,\frac{\Theta_1'(0|is)^{-3}}{\Theta_1(\epsilon|is)}
\,\Theta_1\left(\left.\frac\epsilon2\right|is\right)^4\nonumber\\& &\times
\,\int\limits_{-\infty}^\infty\frac{dq_1}{2\pi}~\e^{-2\pi\alpha'sq_1^2
\left(1+v_1^2\right)-iq_1\left(y_0^{(1)1}-y_0^{(2)1}\right)}\,\delta
\Bigl((v_1-v_2)\beta q_1\Bigr)
\label{freevfinal}\eea
where the impact parameter $b$ of the D$p$-brane scattering is defined through
\be
b^2=\sum_{j=2}^{9-p}\left(y_0^{(1)j}-y_0^{(2)j}\right)^2
\label{impactpar}\ee
In both cases where either $v_1\neq v_2$ or $v_1=v_2$, one reproduces only the
zero temperature results for the scattering amplitude of a pair of moving
D$p$-branes \cite{Bac96,BCD} or the cancellation of the gravitational
attraction with the Ramond-Ramond repulsion between static branes~\cite{Pol95}.

In contrast to that of the charged string, the moving D-brane boundary
state \rf{Dbdry0mode} represents only the ground state configuration. This
difference is easily understood when one considers the Lorentz invariance of
the brane dynamics, i.e. that one can always boost into the rest frame of a
single moving D-brane. A pair of static D$p$-branes corresponds to a neutral
superstring in the T-dual picture, so that it is only possible to write down
a non-trivial temperature dependent interaction between a pair of static
branes \cite{AMS98,Gre92}, as given in section 1. This triviality comes from
the same zero mode operators, associated with the presence of the Wu--Yang
term, as in the electric field problem. It is therefore attributed to the Debye
screening of electric fields that we described in the previous section. In the
present case, the screening comes from the dependence of the static interaction
potential on the gauge field $a_0$ that lives on the brane worldvolume
(see \eq{Tpot}) and it results in a damping of the D-brane motion at finite
temperature. It would be interesting to understand more precisely how the
features of the superstring effective action that we discussed in section 3
affect the non-extremal thermal states of D-branes. In particular, it would be
interesting to understand how the information about a possible deconfinement
phase transition stored in the effective potential for the Polyakov loop
operator is relevant to the corresponding dimensionally-reduced
supersymmetric Yang-Mills theory description at finite temperature. This may
be useful in determining precisely how to unify the two non-extremal
deformations of D-brane configurations (by temperature and velocity) and
thereby describe the thermodynamics of their gravitational interactions. It
may also prove useful to a better understanding of the statistical mechanics
of D-branes.

\subsection*{Acknowledgements}

We are grateful to K.~Zarembo for useful discussions.  The work of J.A., Y.M.
and G.W.S. is supported in part by the grant NATO CRG 970561.
The work of J.A. and Y.M. is supported in part by the grant INTAS 96--0524
and by MaPhySto. The work of Y.M. is supported in part by the grant RFFI
97--02--17927. The work of G.W.S. is supported in part by NSERC of Canada and
the Niels Bohr Fund of Denmark.

\setcounter{section}{0}

\appendix{The Wu--Yang Term \label{AppA}}

In this appendix we will describe the formalism for adding Wu--Yang
terms to the action in the presence of topologically non-trivial gauge fields.
Such terms are generally required whenever a gauge connection on a compact
space has a non-trivial flux. In that case, it is not a globally defined
differential form on the configuration manifold and can only be defined locally
with respect to an open covering of the space. Demanding that the action be
independent of the choice of covering used (or equivalently of the gauge
choice) requires the addition of (generalized) Wu--Yang terms. We shall first
describe the formalism generally in the case of an arbitrary compact manifold
$\cal M$, and then afterwards discuss the specific cases of interest in this
paper. More details can be found in \cite{cech}, for example.

Let $\{U_a\}$ be a finite open cover of the manifold $\cal M$, and assume that
the cover is ``good'', i.e. each $U_a$ and each non-empty intersection of
the $U_a$'s is a contractible open set which is diffeomorphic to an open ball
in ${\bf R}^d$. A gauge field $A$ on a non-trivial line bundle over $\cal M$ is
given by the specification of a one-form $A^{(a)}$ defined everywhere on $U_a$.
On each non-empty intersection of two open sets $U_a$ and $U_b$ of the cover,
the corresponding gauge fields $A^{(a)}$ and $A^{(b)}$ are related by a gauge
transformation:
\be
A^{(a)}-A^{(b)}=d\chi^{(ab)}~~~~~~{\rm on}~~U_a\cap U_b
\label{2overlap}\ee
By definition, the 0-form $\chi^{(ab)}$ satisfies $\chi^{(ab)}=-\chi^{(ba)}$.
Now consider the situation on a non-empty triple overlap of sets $U_a$, $U_b$
and $U_c$. Summing up the three equations of the form (\ref{2overlap}) which
come from the distinct pairwise intersections of the three open sets, we arrive
at the equation
\be
d\left(\chi^{(ab)}+\chi^{(bc)}+\chi^{(ca)}\right)=0
\label{dchi0}\ee
which by Poincar\'e's lemma implies that
\be
\chi^{(ab)}+\chi^{(bc)}+\chi^{(ca)}=c^{(abc)}={\rm const.}~~~~~~{\rm
on}~~U_a\cap U_b\cap U_c
\label{3overlap}\ee
The locally constant functions $c^{(abc)}$ satisfy the cocycle equations
\be
c^{(abc)}-c^{(bcd)}+c^{(cda)}-c^{(dab)}=0
\label{cocycleeq}\ee
and they encode deep topological information
about the line bundle over $\cal M$. They define a two-cocycle of the
\v{C}ech cohomology group $H_{\rm C}^2({\cal M},{\bf R})$ of the manifold $\cal
M$ with coefficients in the constant sheaf $\bf R$. This group measures the
obstructions to passing from local to global data on $\cal M$. There is
a natural isomorphism between the \v{C}ech cohomology group and the ordinary
deRham cohomology group $H_{\rm DR}^2({\cal M})$.

We now consider the appropriate modification of the Wilson loop operator
\be
W[A]=\exp i\oint\limits_\Gamma A
\label{wilsonloop}\ee
integrated over a cycle $\Gamma$ of $\cal M$. Since $A$ is only locally defined
on $\cal M$, this integral needs to be carefully defined using an open cover of
$\cal M$. Consider a triangulation of $\cal M$ such that the simplices induce a
one-dimensional simplicial decomposition of the cycle $\Gamma=\bigcup_aL_a$.
One each line $L_a$ there is a gauge field one-form $A^{(a)}$ as above, and so
naively the appropriate definition of (\ref{wilsonloop}) should be
\be
W[A]=\exp i\sum_a\int\limits_{L_a}A^{(a)}
\label{wilsonmod}\ee
However, the operator (\ref{wilsonmod}) transforms non-trivially under
deformations of the simplicial decomposition. It is straightforward to see that
the induced change in integrand is a sum of terms of the form $d\chi^{(ab)}$
(according to (\ref{2overlap})), so that by Stokes' theorem we should add the
term
$-\sum_{a,b}\chi^{(ab)}(p_{ab})$ to (\ref{wilsonmod}) in order to cancel this
variation. Here $p_{ab}$ is the intersection point of the lines $L_a$ and
$L_b$. We must then cancel out the dependence on the choice of locations of
points $p_{ab}$ in the double overlaps, which according to (\ref{3overlap})
requires the addition of the term $\sum_{a,b,c}c^{(abc)}$.. In this way we
arrive at the consistent topological extension of the Wilson loop integral:
\be
W[A]=\exp
i\left(\sum_a\int\limits_{L_a}A^{(a)}-\sum_{a,b}\chi^{(ab)}(p_{ab})
+\sum_{a,b,c}c^{(abc)}\right)
\label{wilsontopext}\ee
The operator (\ref{wilsontopext}) is independent of the choice of triangulation
of the manifold $\cal M$.

However, the operator (\ref{wilsontopext}) is ambiguous up to the choice of
constants $c^{(abc)}$ (for example, as we will see below, there are choices of
covers which have no triple intersections). Although these constants do not
alter the classical theory, in a path integral approach to the quantum theory
we must demand that the Wilson loop operator (\ref{wilsontopext}) be
independent of the $c^{(abc)}$'s. This imposes the quantization condition
\be
c^{(abc)}=2\pi\,n^{(abc)}~~~~~~,~~~~~~n^{(abc)}\in\bf Z
\label{cocyclequant}\ee
Mathematically, this simply means that $c^{(abc)}/2\pi$ defines a two-cocycle
of the {\it integer} \v{C}ech cohomology $H_{\rm C}^2({\cal M},{\bf Z})$. In
fact, this constraint imposes a quantization condition on the flux of the gauge
field through any two-cycle $\Sigma$ of $\cal M$. To see this we consider the
induced covering of $\Sigma$ by two-simplices $\Delta_a$, such that the
intersection of any two simplices $\Delta_a$ and $\Delta_b$ is a line $L_{ab}$
while the
intersection of any three lines $L_{ab}$, $L_{bc}$ and $L_{ca}$ is a point
$p_{abc}$. We may then compute
\bea
F&\equiv&\frac1{{\rm vol}(\Sigma)}\oint\limits_\Sigma dA~=~\frac1{{\rm
vol}(\Sigma)}\sum_a\int\limits_{\Delta_a}dA^{(a)}\nonumber\\&=&\frac1{{\rm
vol}(\Sigma)}\sum_{a,b}\int\limits_{L_{ab}}\left(A^{(a)}-A^{(b)}
\right)~~~~~~{\rm by~~Stokes'~~theorem}\nonumber\\&=&\frac1{{\rm
vol}(\Sigma)}\sum_{a,b}\int\limits_{L_{ab}}d\chi^{(ab)}~~~~~~{\rm
by~~(\ref{2overlap})}\nonumber\\&=&\frac1{{\rm
vol}(\Sigma)}\sum_{a,b,c}\left(\chi^{(ab)}(p_{abc})
+\chi^{(bc)}(p_{abc})+\chi^{(ca)}(p_{abc})\right)
{}~~~~~~{\rm by~~Stokes'~~theorem}\nonumber\\&=&
\frac1{{\rm vol}(\Sigma)}\sum_{a,b,c}c^{(abc)}
{}~~~~~~{\rm by~~(\ref{3overlap})}
\label{fluxquantcalc}\eea
which using (\ref{cocyclequant}) gives the flux quantization
\be
F=\frac{2\pi N}{{\rm vol}(\Sigma)}
\label{fluxquant}\ee
where $N=\sum_{a,b,c}n^{(abc)}\in\bf Z$.

Let us now turn to the specific examples discussed in the text. For the cases
discussed in sections 2 and 3, we take ${\cal M}={\bf S}^1\times{\bf R}^{d-1}$,
and the Wilson loop integral over the circle ${\bf S}^1$. The minimal good
covering of ${\bf S}^1$ consists of three open sets $U_a$ which
respectively overlie the line segments $[0,\frac\beta3]$,
$[\frac\beta3,\frac{2\beta}3]$ and $[\frac{2\beta}3,\beta]$. The covering may
then be extended trivially through the ${\bf R}^{d-1}$ directions to give a
good cover of the entire manifold $\cal M$. For the gauge choice
(\ref{gauge1}), the transition functions $\chi^{(12)}$ and $\chi^{(23)}$ may be
taken to vanish,
while the third one satisfies (\ref{2overlap}) which gives
\bea
\partial_0\chi^{(13)}&=&0\nonumber\\\vec\partial\chi^{(13)}&=&n\beta c\vec F
\label{chi13}\eea
at $t=0$, where we have used (\ref{pbc}). Integrating (\ref{chi13}) gives
\be
\chi^{(13)}(0)=n\beta c\vec F\cdot\vec x(0)
\label{chi130}\ee
and, since the minimal covering has no triple intersections, the Wilson loop
operator (\ref{wilsontopext}) yields the phase factor (\ref{phase1}). Note
that in this
case there is no analog of the flux quantization condition (\ref{fluxquant})
owing to the absence of non-trivial two-cycles in the present manifold $\cal
M$, or equivalently that $H_{\rm C}^2({\cal M},{\bf Z})=0$ in this case.

Next we consider the case of the manifold ${\cal M}={\bf S}_\beta^1\times
{\bf S}_L^1\times{\bf R}^{d-2}$ which is relevant to the analysis in section 4.
The minimal good cover of the torus ${\bf S}_\beta^1\times{\bf S}_L^1$ consists
of nine open sets $U_a^{(\beta)}\times U_b^{(L)}$ which are obtained from the
product of the minimal good coverings of the circle described above. Again the
covering is trivially extended to the whole of $\cal M$. For the gauge choice
(\ref{gauge1}) it follows from the above example that the only non-vanishing
transition function is $\chi^{(13)}_{\beta L}$ which is induced by those of
${\bf S}_\beta^1$ and ${\bf S}_L^1$. Now, however, the condition
(\ref{2overlap}) reads
\bea
\partial_0\chi_{\beta
L}^{(13)}&=&-mL(1-c)F\nonumber\\\vec\partial\chi^{(13)}_{\beta L}&=&n\beta
c\vec F
\label{chi13bL}\eea
at $t=0$ (with $m$ the winding number around ${\bf S}_L^1$), so that we may
take
\be
\chi_{\beta L}^{(13)}(0)=-mL(1-c)Fx^0(0)+n\beta c\vec F\cdot\vec x(0)
\label{chi13bL0}\ee
The discreteness of the electric field \rf{efieldquant} along the compactified
direction can now be seen as a consequence of the cohomological quantization
condition (\ref{fluxquant}). Taking $\Sigma={\bf S}_\beta^1\times{\bf S}_L^1$,
we have ${\rm vol}(\Sigma)=\beta L$ and thus
\be
F=\frac{2\pi N}{\beta L}~~~~~~,~~~~~~N\in\bf Z
\label{F1quant}\ee
This constraint comes from the mathematical property $H_{\rm C}^2({\cal M},{\bf
Z})=\bf Z$ of the present manifold $\cal M$.

\appendix{Path Integral Evaluation of the Thermal Density Matrix}

To derive \eq{finalrho} from the path integral~\rf{rho}, \rf{phase0},
we use the mode expansion
\be
x^\mu(t)=x^\mu_{\rm cl}(t)+
\sum_{k=1}^\infty \frac{a_k^\mu}{\pi\sqrt{k}}\,\sin \frac{2\pi k t}{s}+
\sum_{k=1}^\infty \frac{b_k^\mu}{\pi \sqrt{k}}
\left(\cos \frac{2\pi k t}{s}-1 \right)
\label{mode}
\ee
where $x^\mu_{\rm cl}(t)$ obeys the classical equations of motion
\bea
\ddot x^0_{\rm cl}+i\vec \E\cdot\dot{\vec x}_{\rm cl} &=& 0 ,\non
\ddot {\vec x}_{\rm cl}-i\vec \E \dot x^0_{\rm cl} &=& 0
\label{claeqn}
\eea
and the boundary conditions~\rf{pbc}. Then the zero mode $x^\mu_{\rm cl}$ is
orthogonal as usual to the modes with $k\geq 1$ with respect to the scalar
product of two functions:
\be
(x,y)= \int\limits_0^s dt~\left( \dot x_\mu \dot y^\mu +
i \E (x^1 \dot y^0 + y^1 \dot x^0)\right) ,
\ee
since the contributions from the non-zero modes vanish
at $t=0$ and $t=s$. Solving \eq{claeqn} we find
\bea
x^0_{\rm cl}&=&y^0+\frac{n\beta}2
\left( 1-\cosh \E t + \coth \frac {\E s}2 \sinh \E t \right)
{}~~~0\leq y^0 <\beta,\non
x^1_{\rm cl}&=&y^1+\frac{in\beta}2
\left( \sinh \E t + \coth \frac {\E s}2
(1 -\cosh \E t )\right),\non
x^i_{\rm cl}&=&y^i ~~~\hbox{for}~~i=2,\ldots,d-1\,.
\label{clasol}
\eea
The factor of $i$ in the second line of \eq{clasol} disappears after the
substitution~\rf{FvsE}.

Substituting~\rf{mode}, \rf{clasol} into the action yields
\be
S=-2\pi \nu n +
S_{\rm cl}+\sum_{k=1}^\infty \sum_{\mu=0}^{d-1}
\frac{k}{s}\,\Bigl(a_k^\mu a_k^\mu + b_k^\mu b_k^\mu \Bigr)
+\sum_{k=1}^\infty \frac{i\E}{\pi}\left(a_k^0 b_k^1-a_k^1 b_k^0 \right)
\ee
with
\be
S_{\rm cl}=\int\limits_0^s dt~
\left( \fr 12  (\dot x^0_{\rm cl})^2+
\fr 12  (\dot x^1_{\rm cl})^2 + i \E x^1_{\rm cl}\dot x^0_{\rm cl}
\right).
\label{Scl}
\ee
The cross term does not appear since the zero modes are orthogonal to the
non-zero ones. To calculate $S_{\rm cl}$ it is convenient to make use of
\eq{claeqn} and an integration by parts in~\rf{Scl} which gives
\be
S_{\rm cl}= \left. \frac 12 x^0_{\rm cl} \dot x^0_{\rm cl} \right\vert_0^s
+\left. \frac 12 x^1_{\rm cl} \dot x^1_{\rm cl} \right\vert_0^s +
\left. \frac {i\E}2 x^1_{\rm cl} \dot x^0_{\rm cl} \right\vert_0^s
=\frac {n^2\beta^2 \E}{4 \tanh \frac {\E s}2} +
i n \beta\E y^1 .
\ee
This results in the argument of the theta function in \eq{finalrho}. The
Gaussian integral over the modes $a_k^0,b_k^0,a_k^1$, and $b_k^1$ with $k\geq1$
produces the fluctuation determinant
\be
\prod_{k=1}^\infty\det \left(
\begin{array}{cccc}
\frac ks & 0&0&-\frac {i\E }{2\pi} \\
0 & \frac ks &\frac {i\E }{2\pi }&0 \\
0&\frac {i\E }{2\pi }& \frac ks &0 \\
-\frac {i\E }{2\pi }&0&0&\frac ks
\end{array}
\right)^{-1/2} = \prod_{k=1}^\infty \left(
\frac{k^2}{s^2} + \frac{\E^2}{4\pi^2}\right)^{-1} =
\frac {\E } {4\pi \sinh \frac{\E s}2} ,
\ee
where we have used zeta-function regularization in the last equality. This
contributes to the pre-exponential factor. The rest of the derivation is
standard.

\end{document}